\documentclass[a4paper]{VisionStyle}
\usepackage{epsfig}

\begin{document}

\title{The Chandra LETG and XMM-Newton Spectra of HR~1099}

\author{M.\,Audard\inst{1} \and M.\,G\"udel\inst{1} \and 
  R.L.J.\,van der Meer\inst{2} } 

\institute{
Paul Scherrer Institut, W\"urenlingen and Villigen, 5232 Villigen PSI, Switzerland
  \and 
Space Research Organization Netherlands, Sorbonnelaan 2, 3584 CA Utrecht, The Netherlands
  }

\maketitle 

\begin{abstract}

The X-ray luminous RS CVn binary system HR~1099 has been 
observed on several occasions in the early phases of \textit{Chandra} and
\textit{XMM-Newton}. A very hot (up to 40~MK) dominant coronal plasma 
has been identified from the high-resolution spectroscopic data;
cooler plasma is seen down to about 3~MK. We recently obtained 
100 ksec  in \textit{Chandra}'s Guest Observer Program to study the corona 
of HR 1099 with the High-Resolution Camera (HRC-S) and the Low 
Energy Transmission Grating (LETG) across the 
complete temperature range above 1~MK.  The data provide an 
unprecedented  view of spectral lines and continua at high 
resolution between ~1 and 175~\AA. We present our investigations
on the \textit{Chandra} LETGS observation of HR~1099 into the 
context of the latest results obtained with \textit{XMM-Newton}.

\end{abstract}

\section{Introduction}
  
\object{HR~1099} (=\object{V711 Tau}; d=28.97~pc, K1~IV+G5~V-IV) is one of 
the brightest binary systems of the RS CVn class.  The active K subgiant mainly
contributes to the chromospheric and coronal emissions (e.g., \cite{maudard-B1-9:ayres01}). 
While in the solar corona, a First Ionization Potential (FIP) effect is observed
(low-FIP elements enhanced by factors of 4-6 relative to high-FIP elements), 
an inverse FIP effect in  the corona of HR~1099 has been found
(\cite{maudard-B1-9:brinkman01}). Other active stars also show such an IFIP effect
(\cite{maudard-B1-9:guedel01a,maudard-B1-9:guedel01b,maudard-B1-9:drake01,maudard-B1-9:huenemoerder01}), 
while it is not present in the intermediately active Capella (\cite{maudard-B1-9:audard01a}). 
It is unclear whether there is an enrichment or a depletion of the coronal
material compared to the photospheric material in RS
CVn systems, because of the high levels of activity and the short rotation
periods. Estimates for HR~1099 range from [Fe/H]$=-0.6$ to 0. Hence the IFIP
effect may only reflect the photospheric composition. However, in an analysis of
solar analogs with known photospheric abundances (close to solar),
\cite*{maudard-B1-9:guedel02} (see also \cite{maudard-B1-9:audard02} in these
proceedings) find an evolution from an IFIP to a normal FIP effect with
decreasing activity, suggesting that the transition is real.

\begin{figure*}
\includegraphics[width=0.5\linewidth]{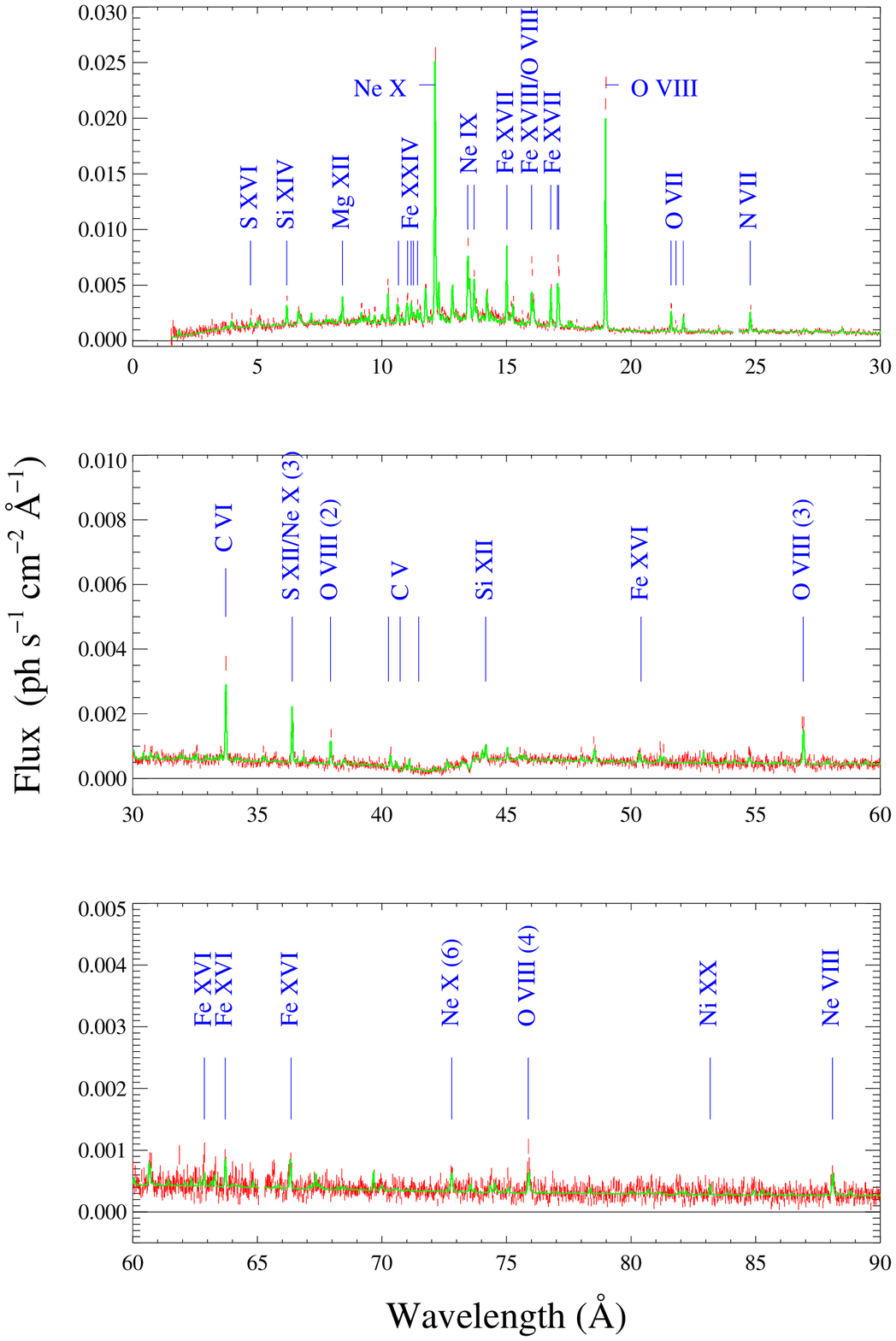}
\hfill
\includegraphics[width=0.5\linewidth]{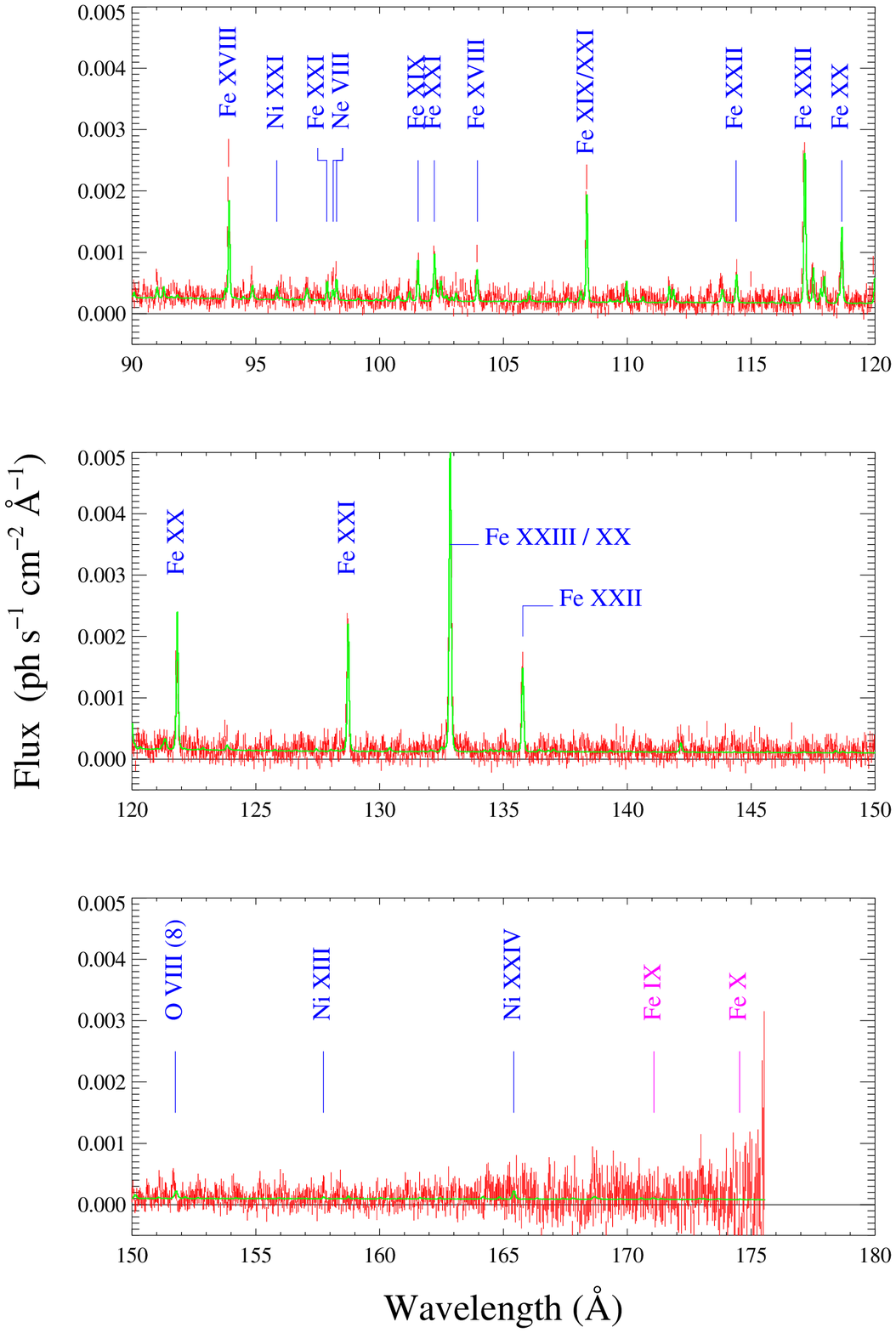}
\vspace*{1mm}
\caption{LETGS spectrum of HR 1099 together with major identified
emission lines (at the exception of the {\rm Fe}~\textsc{ix} and {\rm
Fe}~\textsc{x} lines, in violet, that have been marked to emphasize their 
absence in the spectrum as an evidence of the absence of a significant cool
component, $T \approx 1 - 2$~MK, in the corona of HR~1099). Data are in red, and a best-fit 
model (cf. Fig.~\ref{maudard-B1-9_fig:fig3}) is in green.}
\label{maudard-B1-9_fig:fig1}
\end{figure*}

\section{Analysis} 
\begin{figure}[!ht]
\includegraphics[width=\linewidth]{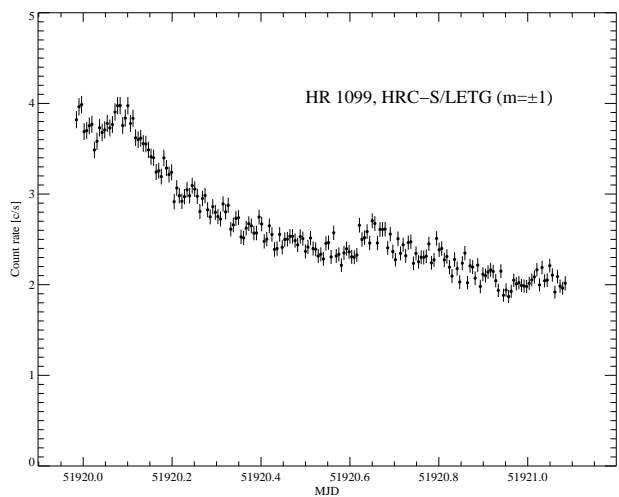}
\caption{X-ray light curve of the LETGS observation of HR~1099. The sum of the
positive and negative grating orders is shown. The light curve suggests
that the observation was possibly performed during the decay phase of a flare.}
\label{maudard-B1-9_fig:fig2}
\end{figure}

\begin{figure}
\centering
\includegraphics[width=\linewidth]{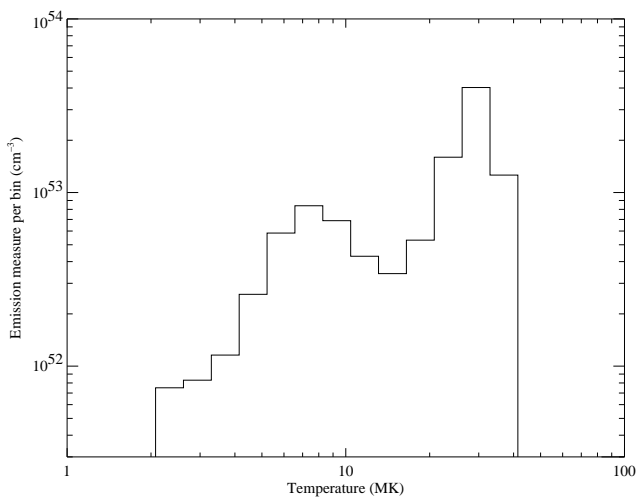}
\caption{Emission measure distribution of HR~1099 observed by
\textit{Chandra} LETGS. Abundances from a multi-temperature fit have
been used. Note that a modified MEKAL plasma emission code available in SPEX
has been used.}
\label{maudard-B1-9_fig:fig3}
\end{figure}

\begin{figure}[!ht]
\centering
\includegraphics[width=\linewidth]{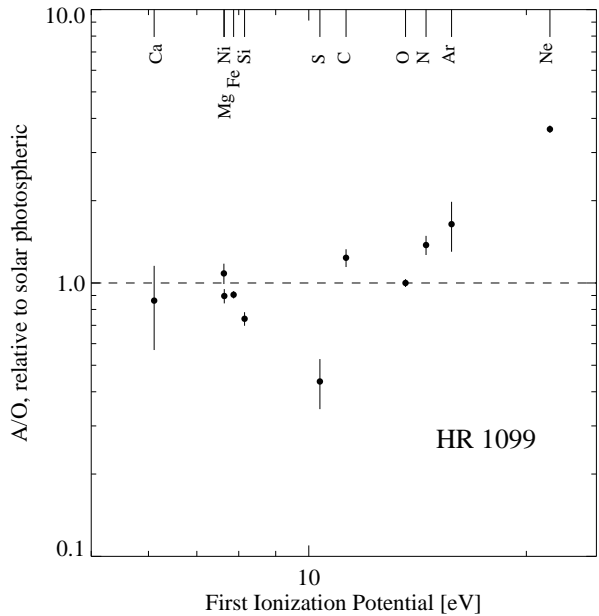}
\vspace*{5mm}
\caption{Coronal abundances of HR~1099 (normalized to oxygen) relative to solar photospheric 
abundances (Anders \& Grevesse 1989, except {\rm Fe}, Grevesse \& Sauval 1999) as a 
function of FIP. Note that the abundances have been derived using a recent
update of the MEKAL code in SPEX. Similarly to Brinkman et al. (2001),
a high Ne/Fe abundance is found, suggesting the presence of an inverse FIP effect.}
\label{maudard-B1-9_fig:fig4}
\end{figure}

\begin{figure}
\centering
\includegraphics[width=\linewidth]{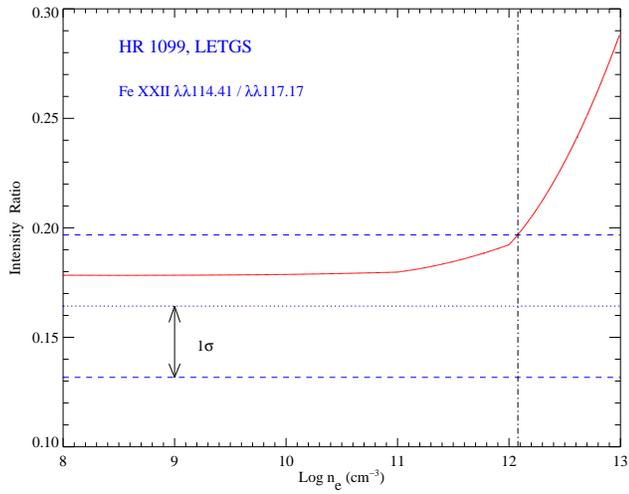}
\caption{Density upper limit from {\rm Fe}~\textsc{xxii} lines. The dotted blue line
represents the measured line ratio, while the dashed blue lines are 1~$\sigma$
confidence ranges. The continuous red curve is a theoretical curve based on
models derived by Brickhouse et al. (1995). The black dash-dotted line gives the
upper limit to the measured density. Note that the {\rm O}~\textsc{vii} 
line ratio (sensitive to plasma $\approx 2$~MK) gives
$n_\mathrm{e} = 2 - 5 \times 10^{10}$~cm$^{-3}$.}
\label{maudard-B1-9_fig:fig5}
\end{figure}

The \textit{Chandra} X-ray Observatory observed \object{HR1099} between January 10--12, 2001 for
95~ksec (obsID1879) with the High-Resolution Camera (HRC-S) and the Low Energy Transmission
Grating (LETG). Its LETGS spectrum displays bright emission lines from H-like 
and He-like transitions of various elements (C, N, O, Ne, Mg, Si, S, Ar, Ca), and a rich forest of lines
from Fe L-shell and 2s-2p lines. A few bright L-shell lines from Si and Ne can 
be identified as well. Figure~\ref{maudard-B1-9_fig:fig1} shows the LETGS 
spectrum from 1.5 to 175\AA\  together with identification labels from major
lines. The X-ray light curve (sum of both grating orders) is also given in
Fig.~\ref{maudard-B1-9_fig:fig2}; the measured flux decreases with time, which
could suggest that the observation has been performed during the decay of a
flare. We have used the updated MEKAL code available in the SPEX fitting
program (\cite{maudard-B1-9:kaastra96}) to fit the average LETGS spectrum (after
co-adding the positive and negative orders). We have performed
a multi-temperature fit to obtain coronal abundances and then used a Chebychev 
polynomial approach to derive the emission measure (EM) distribution. The
latter is dominated by hot plasma ($\approx 20-40$~MK; Fig.~\ref{maudard-B1-9_fig:fig3}).
Coronal abundances (normalized to O) relative to the solar photospheric abundances 
(\cite{maudard-B1-9:anders89}, except for Fe, \cite{maudard-B1-9:grevesse99})
as a function of the First Ionization Potential are shown in
Figure~\ref{maudard-B1-9_fig:fig4}. An enhanced Ne
abundance is found, similarly to results shown by
\cite*{maudard-B1-9:brinkman01}, \cite*{maudard-B1-9:audard01b}, and
\cite*{maudard-B1-9:drake01}. Note that the abundances differ from
those obtained from the \textit{XMM-Newton} RGS data
(see \cite{maudard-B1-9:audard02} in these proceedings), although the same
systematic trend can be found. Most probably, the
discrepancy originates from the use of a different plasma emission code in the
analysis of the RGS data (APEC code). An upper limit of 
$n_\mathrm{e} \leq 10^{12}$~cm$^{-3}$ has been derived from
density-sensitive line ratios of Fe~\textsc{xxii}
(Fig.~\ref{maudard-B1-9_fig:fig5}), while the O~\textsc{vii} line ratio
(sensitive to cool plasma $\approx 2$~MK) gives $n_\mathrm{e}= 2 - 5 \times
10^{10}$~cm$^{-3}$.

\begin{figure*}[!h]
\centering
\includegraphics[width=0.75\linewidth]{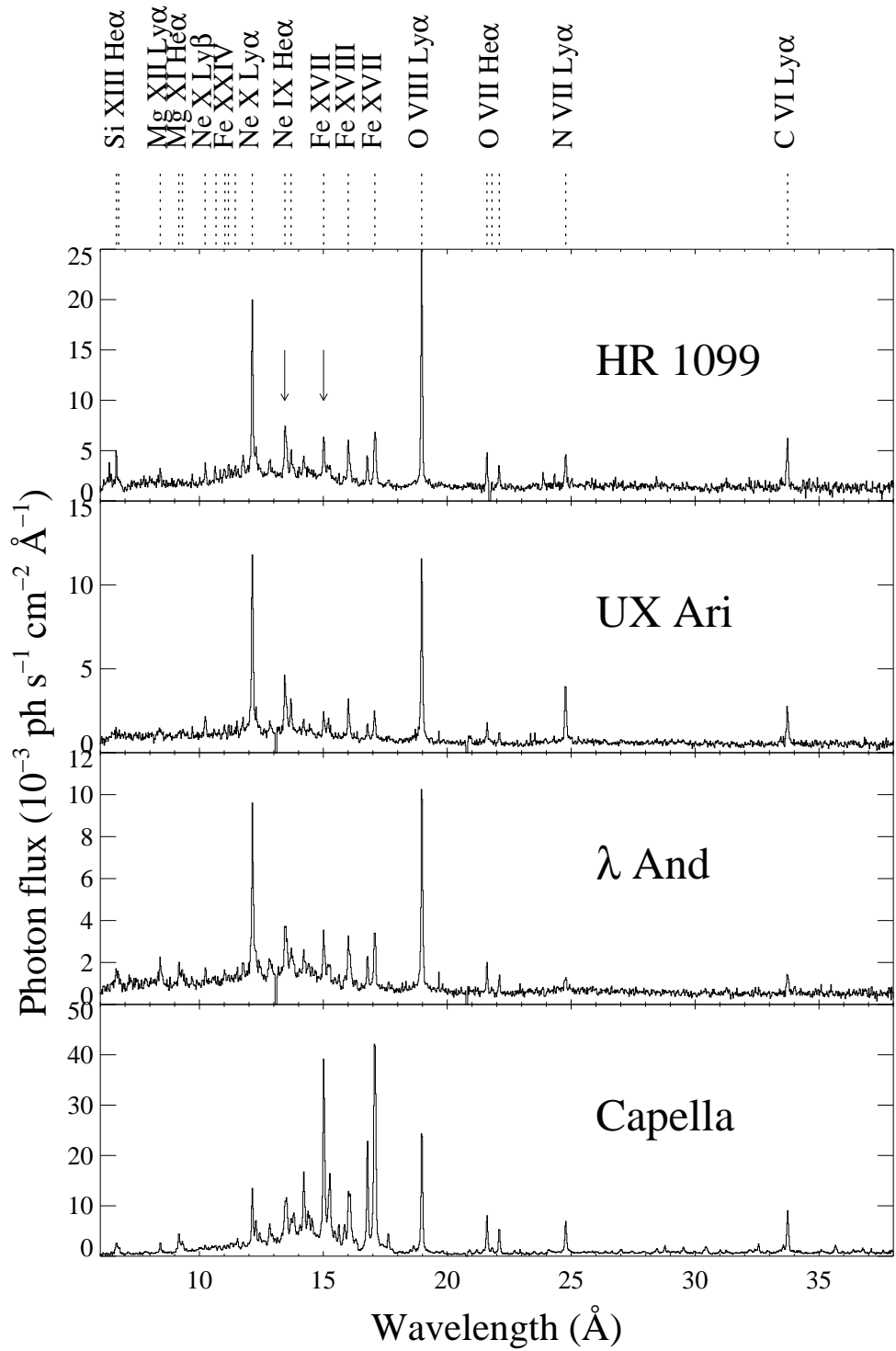}
\caption{\textit{XMM-Newton} RGS 
spectra of HR~1099 with compared other RS CVn binaries (see Audard \& G\"udel 2002 in
these proceedings).  Note the difference in flux ratios between
{\rm Ne}~\textsc{ix} (log$T_{\mathrm{m}} \approx 6.6$; arrow) and 
{\rm Fe}~\textsc{xvii} (log$T_{\mathrm{m}} \approx 6.73$; arrow),
especially in UX Ari and Capella, suggesting that their
coronae differ in their elemental composition.}
\label{maudard-B1-9_fig:fig6}
\end{figure*}

The \textit{XMM-Newton} RGS spectrum of HR~1099 is shown with other RS CVn 
spectra (UX Ari, $\lambda$~And, Capella; Fig.~\ref{maudard-B1-9_fig:fig6}). 
The \textit{Chandra} and
\textit{XMM-Newton} spectra of HR~1099 show a dominance of a hot plasma
indicating a high level of activity in this binary system. As
found by \cite*{maudard-B1-9:brinkman01}, low-FIP elements tend to be
underabundant relative to high-FIP elements. It is probably related
to the correlation between the activity level of a coronal source and its FIP
bias (\cite{maudard-B1-9:guedel02}; see also \cite{maudard-B1-9:audard02}).

\section{Conclusions}
HR~1099 is a hot RS CVn binary that shows a high Ne/Fe ratio. Relative
to solar photospheric abundances, the abundance pattern of HR~1099
(normalized to O) is similar to other active RS CVn binaries, although
less active stars ($\lambda$~And, Capella) show no
inverse FIP effect. This pattern in active binaries fits well
into the long-term evolution from IFIP to FIP found in solar analogs 
(G\"udel et al. 2002; see also \cite{maudard-B1-9:audard02}).

\begin{acknowledgements}

The PSI group acknowledges support from the Swiss National 
Science Foundation (grant 2100-049343). 

\end{acknowledgements}


\begin{thebibliography}{}

\bibitem[\protect\astroncite{Anders \& Grevesse}{1989}]{maudard-B1-9:anders89}
Anders, E., \& Grevesse, N. 1989, Geochim. Cosmoschim. Acta, 53, 197

\bibitem[\protect\astroncite{Audard et al.}{2001a}]{maudard-B1-9:audard01a}
Audard, M., Behar, E., G\"udel, M., et al. 2001a, A\&A, 365, L329

\bibitem[\protect\astroncite{Audard et al.}{2001b}]{maudard-B1-9:audard01b}
Audard, M., G\"udel, M., \& Mewe, R. 2001b, A\&A, 365, L318

\bibitem[\protect\astroncite{Audard \& G\"udel}{2002}]{maudard-B1-9:audard02}
Audard, M., \& G\"udel, M. 2002, these proceedings

\bibitem[\protect\astroncite{Ayres et al.}{2001}]{maudard-B1-9:ayres01}
Ayres, T.~R., Brown, A., Osten, R.~A., et al. 2001, ApJ, 549, 554

\bibitem[\protect\astroncite{Brickhouse et al.}{1995}]{maudard-B1-9:brickhouse95}
Brickhouse, N.~S., Raymond, J.~C., \& Smith, B.~W. 1995, ApJS, 97, 551

\bibitem[\protect\astroncite{Brinkman et al.}{2001}]{maudard-B1-9:brinkman01}
Brinkman, A.~C., Behar, E., G\"udel, M., et al. 2001, A\&A, 365, L324

\bibitem[\protect\astroncite{Drake et al.}{2001}]{maudard-B1-9:drake01}
Drake, J.~J., Brickhouse, N.~S., Kashyap, V.~L., et al. 2001, ApJ, 548, L81

\bibitem[\protect\astroncite{Grevesse \& Sauval}{1999}]{maudard-B1-9:grevesse99}
Grevesse, N., \& Sauval, A.~J. 1999, A\&A, 347, 348

\bibitem[\protect\astroncite{G\"udel et al.}{2001a}]{maudard-B1-9:guedel01a}
G\"udel, M., Audard, M., Briggs, K., et al. 2001a,  A\&A, 365, L336

\bibitem[\protect\astroncite{G\"udel et al.}{2001b}]{maudard-B1-9:guedel01b}
G\"udel, M., Audard, M., Magee, H., et al. 2001b, A\&A, 365, L344

\bibitem[\protect\astroncite{G\"udel et al.}{2002}]{maudard-B1-9:guedel02}
G\"udel, M., Audard, M., Sres, A., et al.
2002, ApJ, submitted

\bibitem[\protect\astroncite{Huenemoerder et al.}{2001}]{maudard-B1-9:huenemoerder01}
Huenemoerder, D.~P., Canizares, C.~R., Schulz, N.~S. 2001, ApJ, 559, 1135

\bibitem[\protect\astroncite{Kaastra, Mewe, \& Nieuwenhuijzen}{1996}]{maudard-B1-9:kaastra96}
Kaastra,  J.~S., Mewe, R., \& Nieuwenhuijzen, H. 1996, in UV and X-ray Spectroscopy of 
 Astrophysical and Laboratory, ed. K. Yamashita \& T. Watanabe (Tokyo: Univ. Acad. Press), 411

\end{thebibliography}
\end{document}